# Existence of a critical layer thickness in PS/PMMA nanolayered films

A. Bironeau[1], T. Salez[2,3], G. Miquelard-Garnier[1,]*, C. Sollogoub[1,]*

*1: PIMM, Arts et Métiers-ParisTech/CNAM/CNRS UMR 8006, 151 bd de l'Hôpital, 75013 Paris, France*

*2: Laboratoire de Physico-Chimie Théorique, UMR CNRS Gulliver 7083, ESPCI Paris, PSL Research University, 75005 Paris, France*

*3: Global Station for Soft Matter, Global Institution for Collaborative Research and Education, Hokkaido University, Sapporo, Hokkaido 060-0808, Japan*

Corresponding authors: cyrille.sollogoub@lecnam.net; guillaume.miquelardgarnier@lecnam.net

**Abstract**

An experimental study was carried out to investigate the existence of a critical layer thickness in nanolayer coextrusion, under which no continuous layer is observed. Polymer films containing thousands of layers of alternating polymers with individual layer thicknesses below 100 nm have been prepared by coextrusion through a series of layer multiplying elements. Different films composed of alternating layers of poly(methyl methacrylate) (PMMA) and polystyrene (PS) were fabricated with the aim to reach individual layer thicknesses as small as possible, varying the number of layers, the mass composition of both components and the final total thickness of the film. Films were characterized by atomic force microscopy (AFM) and a statistical analysis was used to determine the distribution in layer thicknesses and the continuity of layers. For the PS/PMMA nanolayered systems, results point out the existence of a critical layer thickness around 10 nm, below which the layers break up. This critical layer thickness is reached regardless of the processing route, suggesting it might be dependent only on material characteristics but not on process parameters. We propose this breakup phenomenon is due to small interfacial perturbations that are amplified by (van der Waals) disjoining forces.

**for Table of Contents use only**

Existence of a critical layer thickness in PS/PMMA nanolayered films

A. Bironeau, T. Salez, G. Miquelard-Garnier, C. Sollogoub

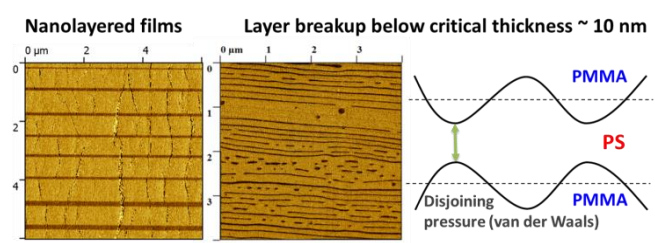

1. Introduction

Nanostructured polymeric materials have shown unique properties arising from the combination of multi-scale assembly, geometrical confinement and interfacial effects.[1–4] The aim of current research activities is thus to develop new strategies to design such nanostructured materials with controlled architecture. In particular, nanolayered (or nanolaminated) structures have received significant attention due to their outstanding mechanical properties observed in natural biological systems like nacre.[5] To fabricate polymer-polymer nanolayered films, different strategies have been reported. A first one, based on a bottom-up approach, consists in non-covalent association of ultra-thin polymer films using molecular self-assembly as a fabrication tool: Langmuir-Blodgett films[6,7] or layer-by-layer assembly.[8,9] However, those techniques suffer mainly from low productivity. Another strategy, that could be assimilated to a top-down approach, consists in using industrial processes, slightly modified or optimized in order to better control the structure down to the nano-scale.[10] One of those structuring processes is the nanolayer coextrusion process, derived from classical coextrusion and capable of producing films with thousands of alternating layers of two polymers A and B, thus yielding individual layer thicknesses down to a few tens of nanometers.[11]

Starting from two or three layers, this process, based on what was termed a forced-assembly concept (as opposed to self-assembly of, for example, block copolymers),[12] has been originally developed in the 1970s by Dow Chemical USA[13] to produce iridescent films industrially. It has then been widely studied by Baer's group over the last twenty years.[11,14] This group has obtained films with improved macroscopic properties (mechanical,[15] optical,[16,17] electrical,[18,19] gas barrier[20,21] …), explained by confinement and/or interfacial effects. The process was also developed in our lab in order to control the architecture at the micro-/nanoscale of multiphase polymer systems, like polymer blends,[22] nanocomposites[23,24] or triblock copolymers.[25]

However, in some cases, it has been observed that below a certain layer thickness, the layers tend to lose their integrity, *i.e.* they break spontaneously during the process. This breakup phenomenon was observed with different polymer pairs and the layer-continuity limit appeared to be system-dependent:

for example, 5-nm thick continuous layers were obtained for polycarbonate (PC) and poly(methyl methacrylate) (PMMA),[26] but nothing thinner than 25 nm has been reported for polypropylene (PP) and polystyrene (PS).[27] Worse, only layers thicker than 500 nm could be achieved for PP and PC.[28]

It is clear that the destruction of the nanolayered structure can have severe consequences and strongly limits the potentiality of this innovative process. In particular, it may alter the final properties, as observed by Lin et al.[29] who have shown a barrier-property loss for polypropylene (PP) / polyethylene oxide (PEO) nanolayer films attributed to layer breakup, occurring when the PEO layer thickness was reduced below 25 nm. It seems therefore of prime importance to better understand the mechanisms governing these layer breakup in order to achieve a well-controlled route towards the design of new nanolayered polymeric materials with enhanced properties. Still, a comprehensive study of the conditions of apparition of these layer breakups at the nanoscale, as well as the physical mechanisms governing them, is lacking in the literature.

Nevertheless, some studies dealing with interfacial distortions or instabilities in coextrusion or the rupture of polymer thin films may shed new light on the nanolayer breakup phenomenon. It may be indeed the consequence of interfacial distortions (viscous encapsulation or secondary flows), mainly encountered when rheologically mismatched polymers are coextruded[30–33] as observed in classical coextrusion. To get rid of these distortions, viscosity and elasticity matching has been a basic rule in coextrusion for a long time.[33,34] Similarly, instabilities initiated by a small perturbation at the interfaces of coextruded polymers that may be eventually amplified along the flow in the die,[35–39] can also induce layer ruptures - especially when the layer thickness is small. If the origin of the initial perturbation is scarcely discussed in the literature, the parameters governing the amplification of the instability have been identified: elastic and viscosity jumps at interfaces .[38,40,41]

Film ruptures quite similar to those obtained in nanolayer coextrusion have been observed by Macosko's group,[42–44] when looking at the morphological development of polymer blends in industrial processes: during the initial stage, softened pellets are stretched and thin polymer sheets are created, that break up through hole formation ("sheeting mechanism"). Still, no precise mechanism is proposed for those film breakups and, in some cases, "Rayleigh instabilities" are erroneously invoked despite their fundamentally different – axisymmetric – origin.[45–47]

Finally, many studies deal with the dewetting of polymer thin films deposited on immiscible polymer substrates in "static" conditions (no shear or elongational flows applied).[48–51] It can be observed for example on thin PS films deposited on PMMA and heated well above the glass-transition temperature $T_g$.[52] Dewetting in thin films is related to interfacial tensions and it spontaneously happens if the spreading parameter defined as $S = \gamma_{PMMA} - (\gamma_{PS} + \gamma_{PS/PMMA})$ is negative. At a molecular level, for films presenting very low thicknesses (< 100 nm), different dewetting routes have been proposed depending on whether the initiation is extrinsic or instrinsic. In the first mechanism, termed nucleation, the presence of nuclei, such as dust particles or surface heterogeneities, triggers topographical defects that will grow into holes. In the second route, called spinodal dewetting by analogy with spinodal decomposition of binary mixtures,[49,53] the mechanism has been proposed by Vrij[54] and Sheludko:[55] thermal fluctuations destabilize the interface and the perturbation can be amplified, if this reduces the system's free energy, leading to the film rupture. Essentially, two ingredients of common – van der Waals – origin are present in this free energy: capillarity, which tends to smoothen and stabilize the interface, and disjoining interactions, acting on distances up to about 100 nm, and which are, depending on the system (nature of the polymer and its environment) either stabilizing or destabilizing.

As a consequence, the aim of the present study is twofold: first, to track the layer breakups when reducing the individual layer thickness in the nanolayered coextrusion process and to determine whether, for a given polymer pair, a critical layer thickness, *i.e.* a thickness below which layers break, can be defined; secondly, to examine and discuss possible mechanisms of layer breakup. The effects of process and material parameters on layer continuity are thus investigated. To avoid crystallization effects and interfacial diffusion, an immiscible glassy polymer pair, PMMA and PS, has been chosen. Films with different processing conditions leading to layer thicknesses ranging from 1 μm down to 2 nm have been produced and characterized.

## 2. Materials and methods

### 2.1. Materials

PMMA was supplied by Altuglas International (Arkema) and is commercially available as Altuglas VM100 (Mass average molar mass $M_w$ = 139 kg.mol$^{-1}$, dispersity $Đ_M$ = 2.1, density at 25 °C = 1.18 g/cm$^3$, density at 200 °C = 1.08 g/cm$^3$). PS, commercially available as Crystal 1340, was provided by Total Petrochemical ($M_w$ = 245 kg.mol$^{-1}$, $Đ_M$ = 2.2, density at 25°C = 1.05 g/cm$^3$, density at 200 °C = 0.96 g/cm$^3$). Molecular weights and dispersities were determined by gel permeation chromatography (GPC) on a Waters 717+ instrument using PS standards for the PS and PMMA standards for the PMMA, with THF (Alfa-Aesar, purity: 99%) as an eluent; density at 25 °C was obtained from the supplier, density at extrusion temperature was measured using a melt-flow indexer (Kayeness) according to ISO 1133. The glass-transition temperatures of PMMA and PS are 95.4 °C and 97.4 °C, respectively, determined by differential scanning calorimetry on a Q10 instrument (TA Instruments). The somewhat low value compared to the typical PMMA value (~105°C) may be due to processing agents added to the polymer grades, since VM100 is an injection grade with low MFI (see below). Nonetheless, the fact that these two values are very close to each other ensures simultaneous shrinkage upon cooling, minimizing the deformation of the multilayered structure.

The melt flow indexes (MFIs) of the polymers studied, as given by the suppliers according to ISO 1133, were 14.5 g/10 min at 230°C/3.8 kg for PMMA VM100 and 4 g/10 min at 200 °C/5 kg for PS 1340. The two polymers have been selected to have a viscosity ratio close to one at the extrusion temperature (225 °C) and in the shear-rate range of the coextrusion process, typically between 1 and 10 s$^{-1}$. Polymer melt rheological properties were measured at 225 °C using an MCR 502 rheometer (Anton Paar) in plate/plate configuration, with a frequency-sweep test (0.01 Hz to 100 Hz at 1% strain). Uniaxial extension tests of selected molten samples were performed using extensional viscosity fixture (EVF, TA Instruments) attached to a strain-controlled rheometer (ARES, TA Instruments). The 18 x 10 x 0.7 mm$^3$ rectangular specimens were prepared by hot-compression molding in the standard mold, provided by TA Instruments with the EVF, at 220 °C and then left under 100 bars for 30 min to relax possible molecular orientation. Selected molten samples were uniaxially extended at 200 °C. Hencky strain rate ($\dot{\varepsilon}$) of 0.1 s$^{-1}$ was applied. All rheological measurements were repeated at least three times for each sample, and their results were averaged.

The obtained values for PMMA and PS lead to a viscosity ratio ($\eta_{PMMA} / \eta_{PS}$) between 0.6 and 0.8 in the relevant shear-rate range. These two materials also showed an elasticity ratio ($G'_{PMMA} / G'_{PS}$) between 0.2 and 0.5, *i.e.* relatively close to 1, as shown by the storage modulus curves. Uniaxial extension tests showed that both polymers have a similar behavior under elongation, typical of linear polymers. Hence, during the process, nanolayered film will be uniformly stretched. All rheological curves are given in the SI. To avoid water uptake and bubbles in the final sample, all products were used under pellet forms and were dried under vacuum for 24h at 80°C prior to processing.

*2.2. Sample preparation*

PMMA/PS nanolayered films are produced using a multilayer coextrusion process. The system is composed of two 20-mm single-screw extruders, two melt-gear pumps, a three-layer (A-B-A) coextrusion feedblock, a layer-multiplying element (LME) assembly, an exit flat die and a thermally regulated chill roll. A schematic illustration is shown in Figure 1. The temperature of feed-block and LME is set to 225 °C. Gear pumps enable a control over the relative composition ratio of the two melt streams that are combined in the A-B-A feedblock. From the feedblock, the initial three-layer flow through a sequence of LME. The melt is initially sliced vertically, and then the halves are spread horizontally to the original width and finally recombined, while keeping the total thickness of the melt constant, hence doubling the number of layers and reducing the thickness of each layer by a factor of 2 after each LME. A series of *N* elements leads to a film composed of $2^{N+1} + 1$ alternating layers, as shown in Figure 1. Here, 10 to 13 LME are used, giving films containing 2049, 4097, 8193 and 16385 layers, respectively. Finally, after passing through the last layer-multiplying element, the melt goes through a flat die, 150 mm wide and 2 mm thick. The exit die temperature is fixed to 200°C. At the die exit, the layered samples were stretched and quenched, using a water-cooled chill roll at a temperature of 95°C, and collected at different drawing speeds. In some cases, two sacrificial polyethylene (PE) skin layers were laminated at the die exit on both sides of the multilayer film, allowing for a reduction of the total film thickness without stretching.

Starting from an A-B-A initial configuration, the expected individual layer thickness of polymer B which will be named nominal thickness ($h_{nom}$) in the manuscript, can be calculated using:

$$h_{nomB} = h_{film} \times \frac{\Phi_B}{n_B} \quad (1),$$

with $h_{film}$ the total film thickness, $\Phi_B$ the volume fraction of polymer B in the film (determined via the weight compositions and densities at extrusion temperature), $n_B = 2^N$ the number of B layers. The equation works similarly for polymer A (with $\Phi_A$, and $n_A = 2^N + 1$).

Looking at Equation (1), it appears that different ways are possible in order to decrease the individual layer thickness: increase the number of LME (which will increase the number of layers without changing the total film thickness), or decrease the total film thickness or relative composition. The draw ratio (Dr) is defined as the roll take-off speed divided by the mean flow speed at the exit die. Hence, increasing Dr and/or adding a skin layer (removed prior to characterization) reduces the total film thickness, *i.e.* decreases the nominal thickness at given number of layers and composition (Dr being inversely proportional to the total thickness of the films). Volume composition is adjusted through the gear-pumps speed. The weight compositions (wt%) of the multilayered PMMA/PS films studied are: 95/5, 90/10, 50/50, 10/90 and 5/95.

The total film thickness ranges from 3000 to 80 μm and the nominal PS and PMMA layer thicknesses were varied from 936 down to 2 nm and from 822 down to 2 nm, respectively. All the PMMA/PS multilayered films investigated in this study can be found in Table S1 (see Supporting Information SI).

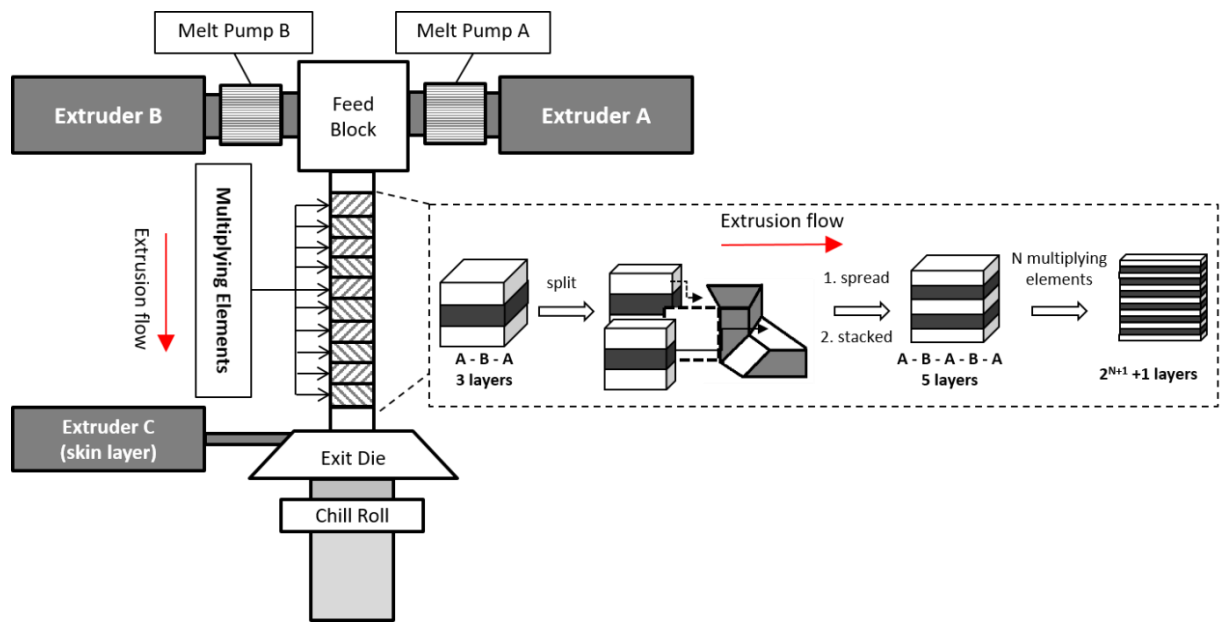

*Figure 1. Principle of the multiplication of layers by the multilayer coextrusion process.*

*2.3. Atomic force microscopy*

Atomic force microscopy (AFM) is used to determine the layer thicknesses, as well as the integrity and uniformity of the films. Samples are cut from the center (along the extrusion axis) of the extruded films and sectioned perpendicular to their surface with an ultramicrotome 2088 UltrotomeV(LKB) at a cutting speed of 1 mm/s. AFM images are obtained in tapping mode using a multimode microscope controlled by a Nanoscope V controller (Veeco), operated under ambient atmosphere. The tips (silicon, spring constant 40 N/m, oscillation frequency 300 kHz) were obtained from BudgetSensors. The curvature radius of the tips is less than 10 nm. A comparative study has been done with a thinner tip (radius of 2 nm) and results regarding layer thicknesses were the same (data not shown), therefore the uncertainty of measurement due to the AFM tip size was considered negligible. Phase, height and amplitude images are acquired simultaneously. AFM images are taken in the extrusion direction (see Figure 2). The layer thicknesses are measured from the AFM phase images that most clearly revealed the layered structure with sharp interfaces. On the obtained images, PS and PMMA appear in brown and gold color, respectively. For all the samples in the study, more than 200 layers were measured.

*2.4. Image analysis*

The layer thicknesses are measured using the AFM phase images and the image analysis software Gwyddion. Through the software, a phase profile can be extracted showing the variation of phase degree. Each layer is represented by one peak on the profile. The thickness of each layer is determined according to a somewhat arbitrary procedure which consists in measuring the full width at half maximum height of the peak, as shown in SI. This measurement method overestimates the value by including the external pixels. Therefore, for each value measured on the profile, the systematic error, *i.e.* the value of one pixel, is subtracted in order to improve accuracy. Based on all the measured thicknesses, it was possible to obtain statistical information which is then used to compare different experimental conditions. The quantities of interest are: the mean thickness ($h_{mean}$) and the thickness distribution (both determined only from the continuous layers) and the percentage of broken layers. The latter is defined as the number of observed broken layers divided by the total number of observed layers.

As layer thickness is expected to be in the range of tens of nanometers, *i.e.* a few pixels in terms of AFM imaging, it is critical to analyze all possible sources of error. These sources of error were studied extensively in a previous article,[56] which we briefly summarize here. The three possible types of error are: uncertainties of measurement, systematic error, and sampling error. The size of the AFM tip, the AFM controller precision, the image compression, and the acquisition definition were considered as uncertainties of measurement. The manual threshold and layer measurement bias due to the operator were considered as systematic error. The sampling which depends on the size of the considered system, *i.e.* the total number of layers to be measured, can be a source of error. The resolution yields a pixel size between 4 and 20 nm depending on the film.

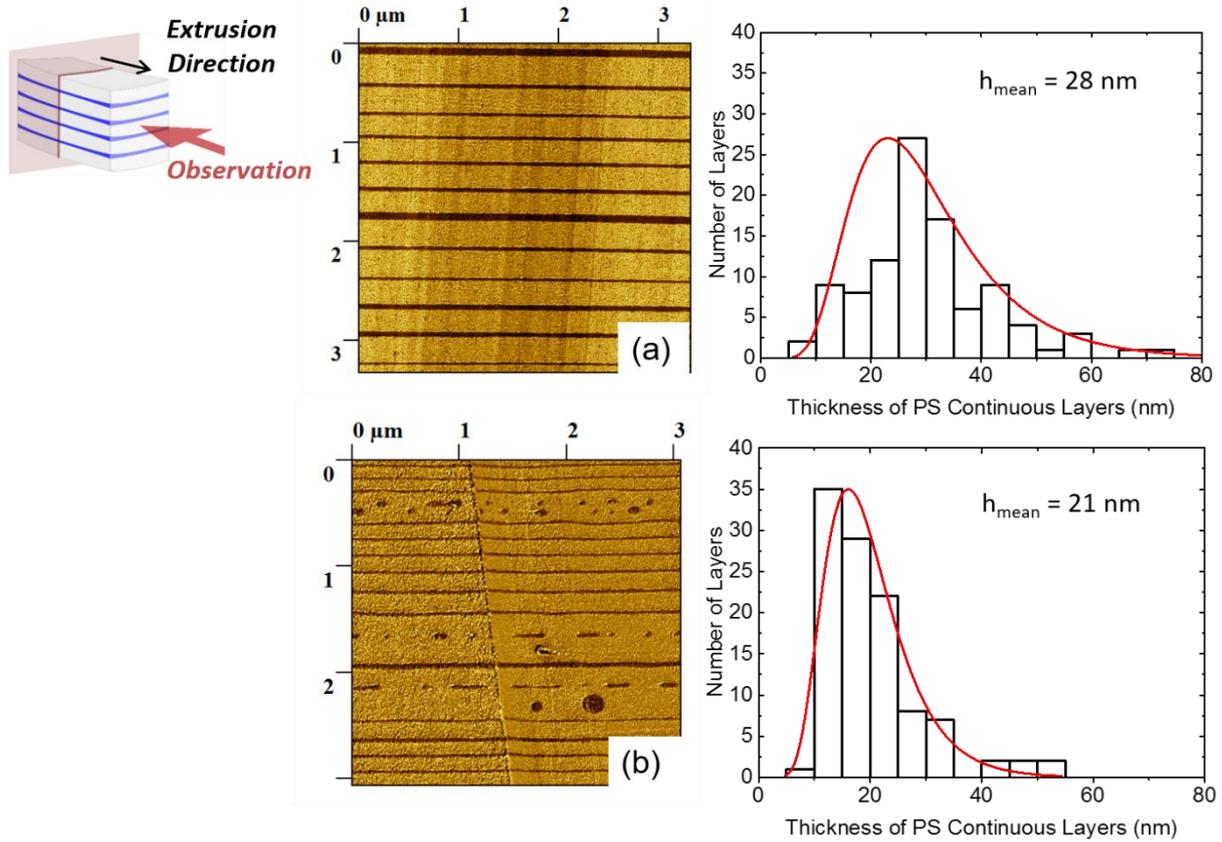

*Figure 2. AFM phase images of PMMA/PS 90/10 wt% nanolayer films with: (a) 27 nm and (b) 22 nm nominal thickness of PS layers; as well as the associated distributions of PS layer thickness. The statistical distribution and in particular $h_{mean}$ are determined only from the continuous layers. The red lines represent the lognormal distribution as a guide to the eye.*

Based on statistical parameters ($K$ a constant and $\gamma$ the scaling-law exponent) determined from a representative-volume-element study for PS layers,[56] and knowing the nominal value $h_{nom}$, the number of performed AFM images, and the size of the images, it was possible to determine the relative uncertainty $\epsilon_{rel}$ of the thickness measurement due to sampling for each film. Calculations have shown that the sampling uncertainty varied between 5% and 30%.

## 3. Results and discussion

### 3.1. Results

Figure 2 shows phase images and statistical distributions of thickness for samples with $h_{nom\ PS}$ = 27 and 24 nm. In both samples displayed in Figure 2, the measured mean thickness (28±18 and 21±10 nm) is quite close to the nominal thickness. For the film with $h_{nom\ PS}$ = 27 nm (Figure 2a), it can be seen that the layers are continuous and the thickness of most of the layers is in the 20–40 nm range. Layers that are observed to be continuous on AFM micrographs are supposed to be continuous all along the sample. However, for the film with $h_{nom\ PS}$ = 22 nm (Figure 2b), disrupted layers and elongated droplets are observed. Similar morphologies have been reported by Liu and coworkers[26] and have been attributed to a surface-tension-driven breakup. For this sample, the percentage of broken PS layers is equal to 22%.

As presented previously, different processing routes are possible in order to achieve a desired final thickness. Figure 3a shows the effects of Dr and composition (weight ratio) on the thickness of the continuous PS layers, keeping the number of layers constant. Conversely, Figure 3b illustrates the effects of Dr and the number of LME keeping the composition constant. As expected, a decrease in layer thickness is observed when decreasing the total film thickness, increasing the number of LME or decreasing the fraction of PS. Figure 3 also shows that for layer thicknesses over 20 nm, the experimentally measured value matches almost perfectly the targeted (nominal) one. However, whatever the processing conditions, the measured thicknesses deviate strongly from the nominal ones, for layer thicknesses below 20 nm. Moreover, no (mean) experimental value below 12 nm is measured, which suggests the existence of a fundamental lower bound for the achievable PS layer thickness obtained via nanolayer coextrusion of PS/PMMA. Same trends were observed for PMMA layers (see Figure S4, SI).

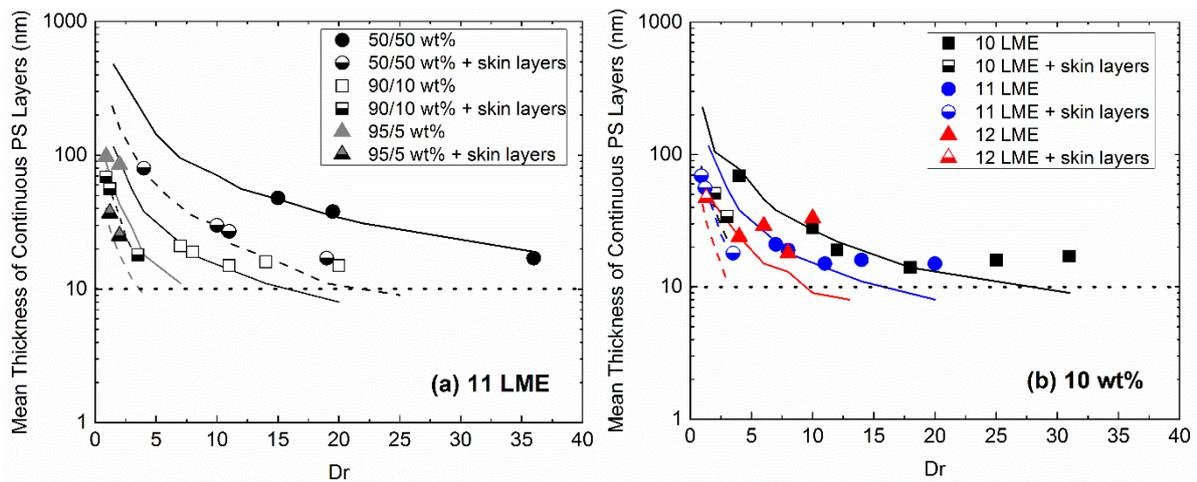

*Figure 3. Mean thickness of continuous PS layers as a function of draw ratio: (a) with 11 LME and at different compositions; (b) with different numbers of LME and at fixed composition (10 wt%). Lines (solid and dashed) correspond to the nominal thickness for the associated conditions, while symbols are measured mean values. Dashed lines indicate the presence of skin layers. The horizontal dotted line indicates a mean thickness of 15 nm. For clarity reasons, the standard deviations are not represented.*

In addition, the percentage of broken PS layers for different Dr is measured and represented in Figure 4a. As stated above, different processing routes can be chosen to reach thicknesses in the 10-nm range: high stretching at the exit die, or high number of LME that may be coupled with a low proportion of the confined polymer and/or the addition of two skin layers. It appears that, regardless of the composition and the number of LME, the layers become more and more discontinuous as Dr is increased, *i.e.* as the film thickness decreases. This result could account for a possible tendency of layers to breakup because of stretching. However, some conditions (high number of LME and/or low volume fraction of one of the polymers) lead to a high percentage (> 50%) of broken layers at low or moderate Dr (grey area in Figure 4a). As a consequence the film stretching induced by the chill roll is not the only step responsible for the layer breakup.

To study more closely the link between the amount of broken layers and the mean thickness, the statistical thickness distribution was built for different Dr. Figure 4b displays this statistical distribution for 10 wt% of PS using 10 LME with Dr ranging from 4 to 31. As already pointed out in Figure 3 for the mean thickness, we observe that the distribution shifts to lower thicknesses when increasing the draw ratio, while the distributions are narrower. As Dr increases from 4 to 18, the mean thickness decreases from 69 nm to 14 nm while the standard deviation decreases from 50 nm to 6 nm (*i.e.* the coefficient of variation decreases from 0.73 to 0.45). We observe as well that the distribution loses its symmetry and becomes truncated at low thickness values. However, at high Dr (Dr =31), the mean thicknesses and standard deviation start to increase again (17 nm and 8 nm respectively). This is correlated to an increase in the percentage of broken layers (see Figure 4a). Those results corroborate the existence of a fundamental critical layer thickness below which layers break up.

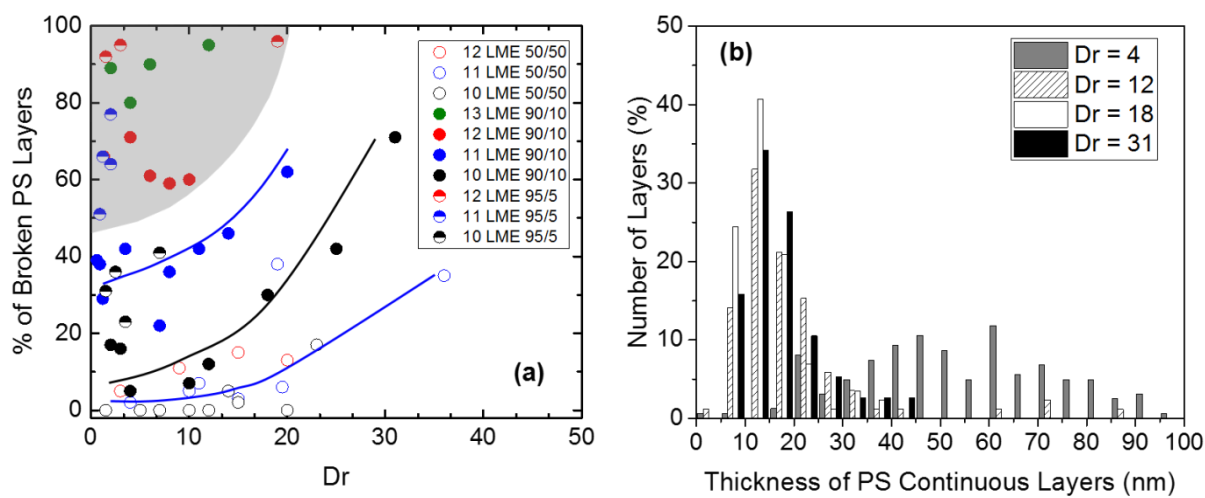

*Figure 4. (a) Percentage of broken PS layers as a function of draw ratio for different numbers of LME and compositions; the color lines are guides to the eye; the grey area indicates a high percentage of broken layers at low draw ratio (b) Distribution of PS layer thickness for different draw ratios for a sample containing 10 wt% of PS, and with 10 LME (corresponding to the black circles in Figure 4a).*

*3.2. Discussion*

Combining all the collected data, two master curves can be plotted as a function of the nominal thickness: the mean experimental thickness (Figure 5a), and the percentage of broken layers (Figure 5b). We chose not to plot the mean thickness when the associated percentage of broken layers is higher than 80%. These master curves allow representing the results for all the processing routes and reveal three distinct regions. In the first one, for nominal layer thicknesses superior to 40 nm, continuous layers are robustly obtained throughout the film (percentage of broken layers lower than 10%) following different processing routes and a good match between nominal and measured layer thicknesses is achieved. In the second one, for nominal layer thicknesses between 10 and 40 nm, all the processing routes are not equivalent and a deviation between the experimental layer thicknesses and the nominal values may occur. Simultaneously, the percentage of broken layers increases. Still, for some optimized processing conditions, the deviations from nominal values remain small and might even be negligible as well as the percentage of broken layers. However, in the third region, for nominal layer thicknesses lower than 10 nm, deviations from the nominal values become significant, the measured value being systematically higher than the nominal one, independently of the processing conditions. These deviations are associated with an important percentage of broken layers, higher than 60% for nominal thickness below 10 nm. Those results confirm once again the existence of a fundamental critical layer thickness below which the layers break up. Specifically, the thinnest layers observed have a thickness of 7 nm (see minimal values plotted in Figure 5a). Looking at a whole sample, no mean thickness lower than 12 nm could be achieved. It is then reasonable to define a critical thickness $h_c$ at around 10 nm for the PS/PMMA system, this critical thickness is obtained independently of the confined polymer, PS or PMMA.

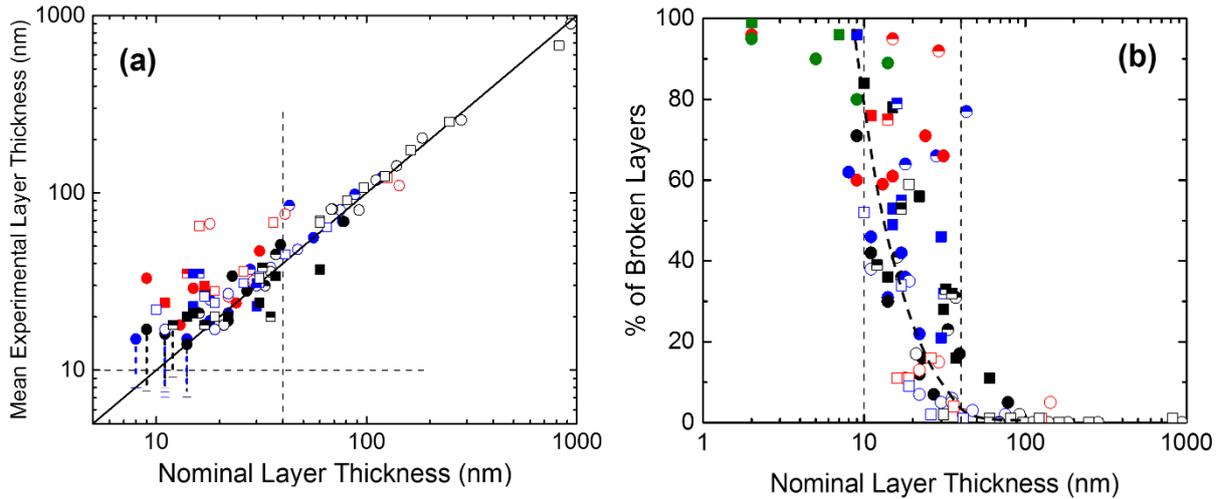

*Figure 5. (a) Mean experimental layer thickness, and (b) percentage of broken layers, as a function of nominal layer thickness, for all processing conditions: PS (circles), PMMA (squares),13 LME (green), 12 LME (red), 11 LME (blue), 10 LME (black), 50/50 wt% (empty), 90/10 wt% (full), 95/5 wt% (half). The thin solid line in Figure 5a is the 1-1 expectation (i.e. $h_{nom} = h_{mean}$), while the thick dashed line in Figure 5b is only a guide to the eye. The regions are delimited by horizontal and vertical thin dashed lines at 10 and 40 nm. The thinnest individual layers measured are indicated through the vertical dotted bars.*

*Critical thickness and possible mechanisms for the layer break-up*

A first basic idea would be that an intrinsic critical thickness should be related to the size of the macromolecules. Indeed, using Kuhn length values from Fetters,[57] one can estimate for the PS used in this study, an average end-to-end radius $R_{PS} \approx 33$ nm, and $R_{PMMA} \approx 23$ nm for PMMA. This is in both cases similar (though slightly bigger) to the observed critical thickness ($h_c \sim 10$ nm). However, this is assuming a random coil conformation, which is certainly not the case for stretched films (Dr > 1) because of the simultaneous drawing and non-uniform cooling of the films at the end of the extruders, leading to different elongated states for the chains among the layers (more elongated near the surface, more relaxed at the center). It should also be noted that stable PS nanolayers much thinner than the radius of gyration can be obtained using other techniques, such as spin-coating, even with higher molecular weights (down to 3 nm,[53] or to 7 nm for stacked spin-coated layers[58]).

Let us then discuss in further details the possible mechanisms mentioned in the introduction for the layer breakup in the nanolayer coextrusion process. Instabilities occurring during classical coextrusion (*i.e.* at the micron scale) have been, as summarized above, widely studied in the literature. In the present study, in order to avoid viscoelastic interfacial distortions or instabilities, rheologically matched PS and PMMA have been chosen (see SI). This rheological matching ensures stable flow and flat (at the microscopic scale) interfaces even for submicronic layers, as it was observed in this study: nearly 0% of broken layers can be achieved for mean layer thicknesses as low as 30 nm with well-chosen experimental conditions (see Figure 5b). This suggests that these mechanisms cannot alone justify what happens for thicknesses below these values.

We then go back to the mechanism responsible for the spinodal dewetting of ultra-thin (< 100 nm) polymer films. In the nanolayered coextruded films, disjoining forces that act on distances up to 100 nm cannot be neglected. When considering two layers of a given polymer (for example PMMA) surrounding a thin layer of another polymer (for example PS), the disjoining forces are attractive and can destabilize the two interfaces.

Following Sheludko,[55] the critical condition for the film rupture can be derived by balancing two opposite forces: the stabilizing capillary force and the destabilizing disjoining force. The disjoining pressure is given by [59]:

$$\pi_{vdW} = \frac{-A_H}{6\pi h^3} \quad (2),$$

where $A_H$ is the Hamaker constant.

The capillary force can be described through the local Laplace pressure developed in the concavity of the disturbed interface:

$$P_c \sim \gamma h'' \sim \frac{\gamma a}{\lambda^2} \quad (3),$$

where $\gamma$ is the interfacial tension, $h$ the thickness of the film, $a$ the amplitude of the instability and $\lambda$ its characteristic wavelength; and where the prime denotes the spatial derivative along one orthogonal direction to the film.

Balancing Equations (2) and (3), and assuming the rupture to occur when $a \sim h_c/2$, we obtain:

$$h_c \sim \left(\frac{A_H \cdot \lambda^2}{3\pi\gamma}\right)^{1/4} \quad (4),$$

The characteristic wavelength $\lambda$ remains an undetermined parameter, but can be chosen as the thickness of the film, as a first estimate. In this case, Equation (4) becomes:

$$h_c \sim \left(\frac{A_H}{3\pi\gamma}\right)^{1/2} \quad (5),$$

A refined approach is to use for $\lambda$ the wavelength of the thermal fluctuations, which can be approximated by $(kT/\gamma)^{1/2}$ [60]. This leads to:

$$h_c \sim \left(\frac{A_H \cdot kT}{3\pi\gamma^2}\right)^{1/4} \quad (5\ bis),$$

In such a mechanism, the layer breakup occurs spontaneously, without any energy barrier. However, it may take a long time, depending on layer thickness: the thicker the films, the weaker the driving force and the longer the breaking time. When the critical thickness is reached, a characteristic rupture time of the film can be derived by balancing the viscous stress and the disjoining pressure:

$$\tau \sim \frac{\eta}{\pi_{vdW}} = \frac{6\pi\eta h_c^3}{A_H} \quad (6),$$

To estimate a critical thickness for our PS/PMMA system from Equation (5), we have to evaluate the Hamaker constant, which is not an easy task, especially in stratified systems where many mutual interactions may have to be considered. Nevertheless, some values can be found in the literature for PS/PMMA bilayer systems deposited on a solid substrate: they cover a few orders of magnitude, between $10^{-18}$ and $10^{-21}$ J, depending on the method used and the environment.[61,62] Considering the value proposed by de Silva et al. for a PS/PMMA/PS trilayer system,[62] $A_{H\ PS/PMMA/PS}$ = $2.10^{-18}$ J (which should be the same value for a PMMA/PS/PMMA system based on Lifshitz theory[59]), we obtain $h_c$ ~14 nm using Equation (5), and $h_c$ ~ 6 nm using Equation (5 bis), which are in good agreement with our experimental findings (Figure 5). If lower values of the Hamaker constant are considered, the critical thickness reaches smaller values, down to ~ 2 nm. For our experimental estimate, $h_c$ ~ 10 nm, the characteristic rupture time calculated from Equation (6) is less than 1 s, which is much less than the processing time (the total mean residence time being around 1 minute). Those estimated critical values confirm that a layer breakup due to interfacial fluctuations amplified by disjoining forces is a realistic scenario in order to explain the experimental results.

Figure 5 shows as well that depending on the processing routes, when the nominal thickness is comprised between 10 and 40 nm, the layer breakup can considerably increase. First, this can be explained by the fact that, when the thickness distribution is large, some layers will reach the critical thickness and consequently break even if the mean thickness is higher than the critical thickness. Secondly, the 10-40 nm thickness range can be considered as a transition region from a capillary-dominated regime to a disjoining-dominated one. Finally, we note that due to the expected presence of impurities in such a semi-industrial process, it is probable that nucleated dewetting occurs at higher thicknesses than the spinodal critical one.

*Comparison with literature data*

The breakup phenomenon in nanolayer coextrusion was also observed with different polymer pairs and appears at different critical thicknesses, as indicated in Table 1. It is important to note that in these previous works no systematic study has been performed in order to ensure that the critical thickness values were independent of the processing conditions. Nevertheless, it is possible to consider such values as critical thicknesses for the considered pairs. Except for PP/PC (the critical thickness value of which may be the result of far-from-optimized processing conditions), all polymer pairs studied lead to similar critical thicknesses, in between 5 and 25 nm.

In our proposed scenario for rupture, the critical thickness is set by the Hamaker constant and the interfacial tension. As stated by Israelachvili,[59] the Hamaker constants of most condensed systems have similar values, and for most polymeric systems should lie in the range $10^{-21}$ - $10^{-18}$ J. Moreover, the values of this constant are not easily found in the literature and can cover the same range for a given pair, as discussed previously for the PS/PMMA system. Similar conclusions can be drawn about most polymer-polymer interfacial tensions: for any given polymer pair, the interfacial tension should lie in the 0.5-5 mJ/m$^2$ range.[63]

Thus, from Equations (6, 6bis) we conclude that the critical thickness should be similar for most polymer pairs and typically close to ~ 10 nm, in agreement with the literature results for amorphous polymeric systems summarized in Table 1. Interestingly, when semi-crystalline polymers are considered instead, the critical thickness appears to be slightly higher.[29] This could reveal the side influence of other phenomena such as volumetric changes during crystallization upon cooling.

A final question is related to the compatibility of the polymer pair and the existence of an interphase (*i.e.* the nanometric region where the polymers are actually blended with each other) which is not accounted for in the proposed mechanism. One way to estimate the compatibility is through the (dimensionless) Flory-Huggins interaction parameter $\chi$, which has been estimated or measured for several polymer pairs. $\chi$ is related to the size $w$ of the interphase and to the interfacial tension, through two equations proposed by Helfand:[64,65]

$$w \approx \frac{2b}{(6\chi)^{1/2}} \quad (7),$$

and:

$$\gamma = \frac{kT}{b^2}\left(\frac{\chi}{6}\right)^{1/2} \quad (8),$$

$b$ being the Kuhn length.[66]

Although these two equations are fairly simple, it should be noted that the determination of the involved parameters, $w$, $\chi$, or $\gamma$ leads to uncertainties since all the quantities are small (a few nanometers for $w$, between $10^{-3}$ and $10^{-1}$ for $\chi$, and usually a few mJ/m$^2$ for $\gamma$),[63] as illustrated in Table 1. Another quantity that encompasses the compatibility between immiscible polymers is the interfacial toughness. As can be seen from partial data that could be obtained in the literature for some polymer pairs (Table 1), no clear trend between these parameters and the critical thickness values can be deduced.

Let us focus on PC/PMMA, one of the most studied polymer pairs in nanolayer coextrusion.[69,12,80] It is considered as a more compatible pair than PS/PMMA, which is apparent when looking at interfacial toughness (more than one order of magnitude of difference) but not when comparing the interaction parameters. Interestingly, slightly lower critical thicknesses have thus been reported for this pair. Moreover, in one of the first studies on this system with a 50/50 percentage volume composition,[12] it was claimed that below nominal layer thicknesses of 12 nm- estimated as the typical size of the interphase by the authors - a new interphase material could be obtained. Blurring of the images was attributed to the interphase having a size similar to the layer thicknesses, thus lowering the contrast. However, in a more recent article, layers having 12 nm nominal thicknesses were shown on AFM images presenting very good contrast.[72] In a subsequent article,[26] layer breakups were observed for nominal thicknesses below 5 nm. These seemingly contradictory conclusions highlight the difficulty of defining a critical thickness when polymer pairs with diffuse interfaces are considered. Therefore, more work is needed in order to achieve a complete understanding of the role of compatibility on the interfacial instabilities occurring in nanolayered polymer flows.

*Table 1. Molecular characteristics and critical layer thickness for polymer-polymer nanolayered coextruded.*

| Polymer pair | Confined polymer | Minimal layer thickness (nm) | Interaction parameter | Interfacial tension (mJ/m$^2$) | Interfacial toughness (J/m$^2$) |
|---|---|---|---|---|---|
| PS / PMMA | PS | 7 | 0.037 at 225 °C[63] | 0.55[67]; 0,56[68] at 225 °C | 45[69] (multilayer); 12[70]; 4-12[71] (multilayer) |
| | PMMA | 10 | | | |
| PP / PC | PP | 500[28] | - | - | - |
| PC / PMMA | PC | 12[72] | 0.039 at 250 °C[73]; 0.017[74] | 1.44 at 240 °C | 1000[75] (multilayer) |
| | PMMA | 5[26] | | | |
| HDPE / PS | HDPE | 10[76] | - | 4[a] | 10[77] |
| PC / PET | PET | 10[78] | - | - | 21[77] |
| PP / PEO | PEO | 25[29] | - | - | - |
| PP / PS | PP | 25[27] | - | 1.4 - 4 at 215 °C[79] | 0[77] |
| | PS | 25[27] | | | |

[a]: extrapolated with linear fit from[68]

## 4. Conclusion

Morphology and layer thicknesses of nanolayered PS/PMMA polymer films processed by coextrusion have been determined through atomic-force microscopy and image analysis. The number of layer multiplying elements, the mass composition, and the total thickness of the films were varied, in order to obtain nanolayered films with nominal layer thicknesses ranging from microns down to a few nanometers. The results revealed that films having nominal layer thicknesses down to 40 nm could be successfully obtained *i.e.* with continuous layers presenting mean thicknesses matching the nominal ones, and no layer breakup. Depending on the processing route, below 40 nm the mean experimental thicknesses appeared to deviate from the nominal ones, along with a substantial increase of the percentage of broken layers. Finally, no film with a mean experimental layer thickness below 10 nm has been obtained. This was interpreted as an evidence for the existence of a fundamental, process-independent critical break-up thickness.

We further suggested the layer breakup phenomenon in the coextrusion process to be due to interfacial instabilities driven by disjoining forces. The thicknesses of the layers we can reach with this process are so small that dispersive forces between two layers composed of the same polymer cannot be neglected (typically below 100 nm). For a thin enough layer, these long-range attractive forces between the surrounding amplify any small disturbance of the interface (*e.g.* induced by thermal fluctuations) leading, after a characteristic time, to the layer breakup. We estimated this characteristic breakup time to be much shorter than the typical processing time. It also appeared that this critical thickness should lie in the same range for most amorphous polymer pairs, assuming crystallization effects or diffuse interfaces can be neglected, which remains an open question.

To further test these hypotheses, simplified experiments on model systems containing a small number of layers can be envisioned. In a recent article,[81] we studied dewetting in a three-layer system (a thin PS film in between two thicker PMMA slabs) heated above both glass-transition temperatures. It was shown that the dewetting kinetics is quite different from what happens in the classical case of substrate-supported thin films, and can be well captured by a simple model balancing capillary and viscous forces. This dewetting kinetics did not depend on the thin-film thickness, and could be seen even on relatively thick films ($h > 200$ nm). We also showed that at temperatures similar to the extrusion temperature of the present study (225°C), dewetting occurs almost instantaneously: within seconds, holes having micrometric diameters could be seen in the PS film. This is very different from what has been reported in the present study. We suggest that the high shear rates induced by extrusion may actually stabilize the layers. This stabilization may be enabled by the lowering of the interfacial thermal-fluctuation amplitude under shear flow, as shown recently by Bickel et al.[82] Similar stabilizations against interfacial instabilities have already been reported, either on Plateau-Rayleigh instabilities in polymer threads[83], or on dewetting in two-layer films.[84] This study based on a process transferable to industry, has raised fundamental questions on polymer thin film stability. Not only could the nanolayer process benefit from this field of research, but it may also be a powerful tool to reassess open questions concerning the physics of polymers under confinement.

**Acknowledgment.** The authors would like to gratefully acknowledge financial support from the French government agency ADEME and Région Aquitaine through the ISOCEL research project. The authors would also like to thank S. Devisme, F. Restagno and J.D. McGraw for fruitful discussions.

**Supporting Information.** Supporting information contains the evolution of the rheological parameters (G', G'', η*) as a function of the angular frequency, the elongational viscosity of the materials used, a complete list of the nanolayered films processed and characterized in this study, a phase profile obtained with the software Gwyddion and the evolution of mean thickness of continuous PMMA layers as a function of draw ratio.